\documentclass[10pt,twocolumn,a4paper]{IEEEtran}
\usepackage{amsmath}
\usepackage{graphicx}

\newcommand{\figuramedia}[3]
{
\begin{figure}
  \centering
 \includegraphics[width=6cm]{#1}
  \caption{#2}\label{#3}
\end{figure}
}

\newcommand{\figura}[3]
{
\begin{figure}
  \centering
 \includegraphics[width=8cm]{#1}
  \caption{#2}\label{#3}
\end{figure}
}
\title{Unsaturated Throughput Analysis of IEEE 802.11 in Presence of Non Ideal Transmission
Channel and Capture Effects}
\author{\authorblockN{F. Daneshgaran, M. Laddomada, F. Mesiti, and M.
Mondin\thanks{This work has been partially supported by
Euroconcepts, S.r.l. (http://www.euroconcepts.it) and ISMB through
the European IST project, OBAN
(http://oban.prz.tu-berlin.de/index.html), IST 6FP Contract No.
001889.}
\thanks{\textbf{To appear on IEEE Transactions on Wireless Communications, 2008}}
\thanks{F. Daneshgaran is with Euroconcepts,
S.r.l., Italy.}
\thanks{M. Laddomada (\textrm{laddomada@polito.it}), F. Mesiti and M. Mondin
are with DELEN, Politecnico di Torino, Italy.}}}
\begin{document}
\maketitle

\begin{abstract}
In this paper, we provide a throughput analysis of the IEEE 802.11
protocol at the data link layer in non-saturated traffic
conditions taking into account the impact of both transmission
channel and capture effects in Rayleigh fading environment. The
impact of both non-ideal channel and capture become important in
terms of the actual observed throughput in typical network
conditions whereby traffic is mainly unsaturated, especially in an
environment of high interference.

We extend the multi-dimensional Markovian state transition model
characterizing the behavior at the MAC layer by including
transmission states that account for packet transmission failures
due to errors caused by propagation through the channel, along
with a state characterizing the system when there are no packets
to be transmitted in the buffer of a station.
Finally, we derive a linear model of the throughput
along with its interval of validity.

Simulation results closely match the theoretical derivations
confirming the effectiveness of the proposed model.
\end{abstract}
\section{Introduction}
\IEEEPARstart{W}{ireless} Local Area Networks (WLANs) using the
IEEE802.11 series of standards have experienced an exponential
growth in the recent past [1-23]. The Medium Access Control (MAC)
layer of many wireless protocols resemble that of IEEE802.11.
Hence, while we focus on this protocol, it is evident that the
results easily extend to other protocols with similar MAC layer
operation.

The IEEE802.11 MAC presents two options \cite{standard_DCF_MAC},
namely the Distributed Coordination Function (DCF) and the Point
Coordination Function (PCF). PCF is generally a complex access
method that can be implemented in an infrastructure network. DCF
is similar to Carrier Sense Multiple Access with Collision
Avoidance (CSMA/CA) and is the focus of this paper.

With this background, let us provide a quick survey of the recent
literature related to the problem addressed here. This survey is
by no means exhaustive and is meant to simply provide a sampling
of the literature in this fertile area. We invite the interested
readers to refer to the references we provide in the following and
references therein.

The most relevant works to what is presented here are
\cite{Bianchi,Liaw,Ergen}. In~\cite{Bianchi} the author provided
an analysis of the saturation throughput of the basic 802.11
protocol assuming a two dimensional Markov model at the MAC layer,
while in~\cite{Liaw} the authors extended the underlying model in
order to consider unsaturated traffic conditions by introducing a
new idle state, not present in the original Bianchi's model,
accounting for the case in which the station buffer is empty,
after a successful completion of a packet transmission. In the
modified model, however, a packet is discarded after $m$ backoff
stages, while in Bianchi's model, the station keeps iterating in
the $m$-th backoff stage until the packet gets successfully
transmitted.
In~\cite{Ergen}, authors propose a novel Markov model for the DCF
of IEEE 802.11, using a IEEE 802.11a PHY, in a scenario with
various stations contending for the channel and transmitting with
different transmission rates. An admission control mechanism is
also proposed for maximizing the throughput while guaranteeing
fairness to the involved transmitting stations.

In \cite{Randhawa}, the authors extend the work of Bianchi to
multiple queues with different contention characteristics in the
802.11e variant of the standard with provisions of QoS. In
\cite{kong}, the authors present an analytical model, in which
most new features of the Enhanced Distributed Channel Access
(EDCA) in 802.11e such as virtual collision, different arbitration
interframe spaces (AIFS), and different contention windows are taken
into account. Based on the model, the throughput performance of
differentiated service traffic is analyzed and a recursive method
capable of calculating the mean access delay is presented. Both
articles referenced assume the transmission channel to be ideal.

In \cite{Qiao}, the authors look at the impact of channel induced
errors and the received SNR on the achievable throughput in a
system with rate adaptation whereby the transmission rate of the
terminal is adapted based on either direct or indirect
measurements of the link quality. In \cite{Chatzimisios}, the
authors deal with the extension of Bianchi's Markov model in order
to account for channel errors.
In \cite{QiangNi}, the authors
investigate the saturation throughput in both congested and
error-prone Gaussian channel, by proposing a simple and accurate
analytical model of the behaviour of the DCF at the MAC layer. The
PHY-layer is based on the parameters of the IEEE 802.11a
protocol.

Paper~\cite{Malone} proposes an extension of the Bianchi's model
considering a new state for each backoff stage accounting for the
absence of new packets to be transmitted, i.e., in unloaded
traffic conditions.

\noindent In real networks, traffic is mostly unsaturated, so it
is important to derive a model accounting for practical network
operations. In this paper, we extend the previous works on the
subject by looking at all the three issues outlined before
together, namely real channel conditions, unsaturated traffic, and
capture effects. Our assumptions are essentially similar to those
of Bianchi~\cite{Bianchi} with the exception that we do assume the
presence of both channel induced errors and capture effects due to
the transmission over a Rayleigh fading channel. As a reference
standard, we use network parameters belonging to the IEEE802.11b
protocol, even though the proposed mathematical models hold for
any flavor of the IEEE802.11 family or other wireless protocols
with similar MAC layer functionality.
We also demonstrate
that the behavior of the throughput versus the packet rate,
$\lambda$, is a linear function of $\lambda$ up to a critical
packet rate, $\lambda_c$, above which throughput enters saturated
conditions. Furthermore, in the linear region, the slope of the
throughput depends only on the average packet length and on the
number of contending stations.

Paper outline is as follows. After a brief review of the
functionalities of the contention window procedure at MAC layer,
section~II extends the Markov model initially proposed by Bianchi,
presenting modifications that account for transmission errors and
capture effects over Rayleigh fading channels employing the 2-way
handshaking technique in unsaturated traffic conditions.
Section~III provides an expression for the unsaturated throughput
of the link. In section~IV we present simulation results where
typical MAC layer parameters for IEEE802.11b are used to obtain
throughput values as a function of various system level
parameters, capture probability, and SNR under typical traffic
conditions. Section~V derives a linear model of the throughput
along with its interval of validity. Finally, Section~VI is
devoted to conclusions.
\section{Development of the Markov Model}
In this section, we present the basic rationales of the proposed
bi-dimensional Markov model useful for evaluating the throughput
of the DCF under unsaturated traffic conditions. We make the
following assumptions: the number of terminals is finite, channel is prone to errors,
capture in Rayleigh fading transmission
scenario can occur, and packet transmission is based on the 2-way handshaking
access mechanism. For conciseness, we will limit our presentation
to the ideas needed for developing the proposed model. The
interested readers can refer to \cite{standard_DCF_MAC,Bianchi}
for many details on the operating functionalities of the DCF.
\subsection{Markovian Model Characterizing the MAC Layer under unsaturated traffic conditions,
Real Transmission Channel and Capture Effects}
In \cite{Bianchi}, an analytical model is presented for the
computation of the throughput of a WLAN using the IEEE 802.11 DCF
under ideal channel conditions. By virtue of the strategy employed
for reducing the collision probability of the packets transmitted
from the stations attempting to access the channel simultaneously,
a random process $b(t)$ is used to represent the backoff counter
of a given station. Backoff counter is decremented at the start of
every idle backoff slot and when it reaches zero, the station
transmits and a new value for $b(t)$ is set.

The value of $b(t)$ after each transmission depends on the size of
the contention window from which it is drawn. Therefore it depends
on the stations' transmission history, rendering it a
non-Markovian process. To overcome this problem and get to the
definition of a Markovian process, a second process $s(t)$ is
defined representing the size of the contention window from which
$b(t)$ is drawn, $(W_i=2^i W,~i=s(t))$.

We recall that a backoff time counter is initialized depending on
the number of failed transmissions for the transmitted packet. It
is chosen in the range $[0,W_i-1]$ following a uniform
distribution, where $W_i$ is the contention window at the backoff
stage $i$. At the first transmission attempt (i.e., for $i=0$),
the contention window size is set equal to a minimum value
$W_0=W$, and the process $s(t)$ takes on the value $s(t)=i=0$.

The backoff stage $i$ is incremented in unitary steps after each
unsuccessful transmission up to the maximum value $m$, while the
contention window is doubled at each stage up to the maximum value
$CW_{max}=2^m W$.

The backoff time counter is decremented as long as the channel is
sensed idle and stopped when a transmission is detected. The
station transmits when the backoff time counter reaches zero.

A two-dimensional Markov process $(s(t),b(t))$ can now be defined,
based on two assertions:
\begin{enumerate}
\item the probability $\tau$ that a station will attempt
transmission in a generic time slot is constant across all time
slots;
\item the probability $P_{col}$ that any transmission experiences
a collision is constant and independent of the number of
collisions already suffered.
\end{enumerate}
Bianchi's model relies on the following fundamental assumptions:
1) the mobile stations always have something to transmit (i.e.,
the saturation condition), 2) there are no hidden terminals and
there is no capture effect (i.e., a terminal which perceives a
higher signal-to-noise ratio (SNR) relative to other terminals
capturing the channel~\cite{capture,zorzi_rao,Spasenovski} and
limiting access to other terminals, similar to the near-far
problem in cellular networks), 3) at each transmission attempt and
regardless of the number of retransmissions suffered, each packet
collides with constant and independent probability, and 4) the
transmission channel is ideal and packet errors are only due to
collisions. Clearly, the first, the second and fourth assumptions
are not valid in any real setting, specially when there is
mobility and when the transmission channel suffers from fading
effects.

The main aim of this section is to propose an effective
modification of the bi-dimensional Markov process proposed
in~\cite{Bianchi} in order to account for unsaturated traffic
conditions, channel error propagation and capture effects over a
Rayleigh fading channel under the hypothesis of employing a 2-way
handshaking access mechanism.

It is useful to briefly recall the
2-way handshaking mechanism.
A station that wants to transmit a packet, waits until the channel
is sensed idle for a Distributed InterFrame Space (DIFS), follows
the backoff rules and then transmits a packet. After the
successful reception of a data frame, the receiver sends an ACK
frame to the transmitter. Only upon a correct ACK frame reception,
the transmitter assumes successful delivery of the corresponding
data frame. If an ACK frame is received in error or if no ACK
frame is received, due possibly to an erroneous reception of the
preceding data frame, the transmitter will contend again for the
medium.

On the basis of this assumption, collisions can occur with
probability $P_{col}$ on the transmitted packets, while
transmission errors due to the channel, can occur with
probability $P_e$. We assume that collisions and transmission error
events are statistically independent. In this scenario, a packet
is successfully transmitted if there is no collision (this event
has probability $1-P_{col}$) and the packet encounters no channel
errors during transmission (this event has probability $1-P_{e}$).
The probability of successful transmission is therefore equal to
$(1-P_{e})(1-P_{col})$, from which we can set an equivalent
probability of failed transmission as $P_{eq}=P_e+P_{col}-P_e
P_{col}$.

Furthermore, in mobile radio environment, it may happen that the
channel is captured by a station whose power level is stronger
than other stations transmitting at the same time. This may be due
to relative distances and/or channel conditions for each user and
may happen whether or not the terminals exercise power control.
Capture effect often reduces the collision probability on the
channel since the stations whose power level at the receiver are
very low due to path attenuation, shadowing and fading, are
considered as interferers at the access point (AP) raising the
noise floor.
\figuramedia{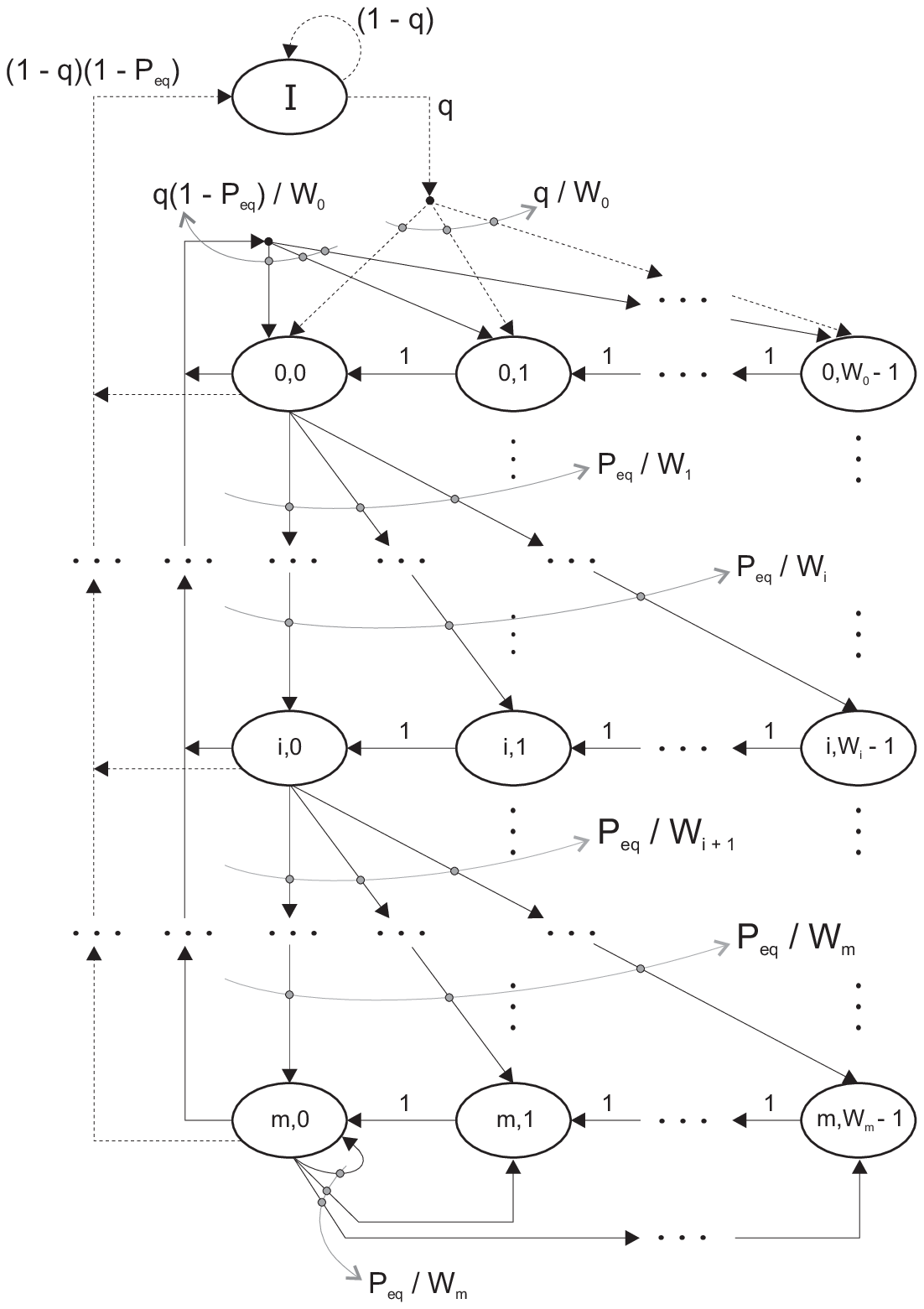}{Markov chain for
the contention model in unsaturated traffic conditions, based on
the 2-way handshaking technique, considering the effects of
capture and channel induced errors.}{fig.chain}

To simplify the analysis, we make the assumption that the impact
of the channel induced errors on the packet headers are negligible
because of their short length with respect to the data payload
size. This is justified on the basis of the assumption that the
bit errors affecting the transmitted data are independent of each
other. Hence, the packet or frame error rate, identified
respectively with the acronyms PER or FER, is a function of the
packet length, with shorter packets having exponentially smaller
probability of error compared to longer packets.

Practical networks operate in unsaturated traffic conditions. In
this case, Bianchi's model~\cite{Bianchi} assuming the presence of
a packet to be transmitted in each and every station's buffer, is not valid
anymore. However, the simplicity of such a model can be retained
also in unsaturated conditions by introducing a new state,
labelled $I$, accounting for the following two situations:
\begin{itemize}
    \item immediately after a successful transmission, the buffer of the
    transmitting station is empty;
    \item the station is in an idle state with an empty buffer until a new
    packet arrives at the buffer for transmission.
\end{itemize}
With these considerations in mind, let us discuss the Markov model
shown in Fig.~\ref{fig.chain}, modelling unsaturated traffic
condition.
Notice that in our Markov chain we do not model
the 802.11 post-backoff feature. However, post-backoff has
negligible effects on the theoretical results developed in the paper
as exemplified by the simulation results obtained in the
following.
Similar to the model in~\cite{Bianchi}, $(m+1)$
different backoff stages are considered (this includes the zero-th
stage). The maximum contention windows (CW) size is $CWmax =
2^mW$, and the notation $W_i = 2^iW$ is used to define the
$i^{th}$ contention window size. A packet transmission is
attempted only in the $(i,0)$ states, $\forall i=0,\ldots,m$. If
collision occurs, or transmission is unsuccessful due to channel
errors, the backoff stage is incremented, so that the new state
can be $(i+1,k)$ with probability $P_{eq}/W_{i+1}$, since a
uniform distribution between the states in the same backoff stage
is assumed. We consider capture as a subset of the event of a
collision (i.e., capture implicitly implies a collision). In other
word, the capture event can happen in the presence of collision by
allowing the transmitting station with the highest received power
level at the access point to capture the channel. In the event of
capture, no collision is detected and the Markov model transits
into one of the transmitting states $(i,0)$ depending on the
current contention stage. If no collision occurs (or is detected),
a data frame can be transmitted and the transmitting station
enters state $(i,0)$ based on its backoff stage. From state
$(i,0)$ the transmitting station re-enters the initial backoff
stage $i=0$ if the transmission is successful and at least one
packet is present in the buffer, otherwise the station transit in
the state labelled $I$ waiting for a new packet arrival.
Otherwise, if errors occur during transmission, the ACK packet is
not sent, an ACK-timeout occurs, and the backoff stage is changed
to $(i+1,k)$ with probability $P_{eq}/W_{i+1}$.

The Markov Process of Fig.~\ref{fig.chain} is governed by the
following transition probabilities\footnote{$P_{i,k|j,n}$ is short
for $P\{s(t+1)=i,b(t+1)=k|s(t)=j,b(t)=n\}$.}:
\begin{equation}\label{eq.process}\small
\begin{array}{lll}
P_{i,k|i,k+1} &= 1,                     &~ k \in [0,W_i-2], ~ i
\in [0,m] \\
P_{0,k|i,0}  &= q(1-P_{eq})/W_0, &~ k \in [0,W_0-1], ~ i \in [0,m]
\\ P_{i,k|i-1,0}   &= P_{eq}/W_i,&~ k \in [0,W_i-1], ~ i \in [1,m]
\\ P_{m,k|m,0}   &= P_{eq}/W_m,&~ k \in [0,W_m-1]\\
P_{I|i,0}  &= (1-q)(1-P_{eq}), &~ i \in [0,m]\\
P_{0,k|I}  &= q/W_0, &~ k \in [0,W_0-1] \\
P_{I|I}  &= 1-q &
\end{array}
\end{equation}
The first equation in~(\ref{eq.process}) states that, at the
beginning of each slot time, the backoff time is decremented. The
second equation accounts for the fact that after a successful
transmission, a new packet transmission starts with backoff stage
0 with probability $q$, in case there is a new packet in the
buffer to be transmitted. Third and fourth equations deal with
unsuccessful transmissions and the need to reschedule a new
contention stage. The fifth equation deals with the practical
situation in which after a successful transmission, the buffer of
the station is empty, and as a consequence, the station transits
in the idle state $I$ waiting for a new packet arrival. The sixth
equation models the situation in which a new packet arrives in the
station buffer, and a new backoff procedure is scheduled. Finally,
the seventh equation models the situation in which there are no
packets to be transmitted and the station is in the idle state.
\section{Markovian Process Analysis and Throughput Computation}
Next line of pursuit consists in finding a solution of
the stationary distribution:
\[
b_{i,k}=\lim_{t\rightarrow \infty}P[s(t)=i,b(t)=k],~\forall
k\in[0,W_i-1],~\forall i\in[0,m]
\]
that is, the probability of a station occupying a given state at
any discrete time slot.

First, we note the following relations:
\begin{equation}\label{trans_states_probabilities}
\begin{array}{rcll}
b_{i,0} & = & P_{eq} \cdot b_{i-1,0} = P_{eq}^i \cdot b_{0,0},   &\forall i \in [1,m-1] \\
b_{m,0} & = & \frac{P_{eq}^m}{1-P_{eq}} \cdot b_{0,0},           & i = m \\
\end{array}
\end{equation}
whereby, $P_{eq}$ \footnote{For simplicity, we assume that at each
transmission attempt any station will encounter a constant and
independent probability of failed transmission, $P_{eq}$,
independently from the number of retransmissions already suffered
from each station.} is the equivalent probability of failed
transmission, that takes into account the need for a new
contention due to either packet collision ($P_{col}$) or channel
errors ($P_e$), i.e.,
\begin{equation}\label{eq.equ}
P_{eq}=P_{col}+P_e-P_e\cdot  P_{col}
\end{equation}
State $b_I$ in Fig.~\ref{fig.chain} considers both the situation
in which after a successful transmission there are no packets to
be transmitted, and the situation in which the packet queue is
empty and the station is waiting for new packet arrival.
The stationary probability to be in state $b_I$ can
be evaluated as follows:
\begin{equation}
\label{eq:b_N}
\begin{array}{rcl}
b_I & = & (1-q)(1-P_{eq}) \cdot \sum_{i=0}^{m}b_{i,0} + (1-q) \cdot b_I \\
    & = & \frac{(1-q)(1-P_{eq})}{q} \cdot \sum_{i=0}^{m}b_{i,0} \\
\end{array}
\end{equation}
The expression above reflects the fact that state $b_I$ can be reached after a successful packet
transmission from any state $b_{i,0},~\forall i \in [0,m]$ with
probability $(1-q)(1-P_{eq})$, or because the station is waiting
in idle state with probability $(1-q)$,
whereby $q$ is the probability of having at least one packet to be
transmitted in the buffer. The statistical model of $q$ will be
discussed in the next section.

The other stationary probabilities for any $k\in[1,W_i-1]$ follow
by resorting to the state transition diagram shown in
Fig.~\ref{fig.chain}:
\begin{equation}\label{eq.bik}
b_{i,k} = \frac{W_i-k}{W_i} \left\{
\begin{array}{ll}
q(1-P_{eq}) \cdot \sum_{i=0}^{m}b_{i,0} + q \cdot b_I,       & i = 0 \\
P_{eq} \cdot b_{i-1,0},                                      & i \in [1,m-1] \\
P_{eq} (b_{m-1,0} + b_{m,0}),                                & i = m \\
\end{array}\right.
\end{equation}
Upon substituting~(\ref{eq:b_N}) in~(\ref{eq.bik}), $b_{0,k}$ can
be rewritten as follows:
\begin{equation}
\begin{array}{ll}
 q(1-P_{eq}) \cdot \sum_{i=0}^{m}b_{i,0} + q \cdot b_I =&\\
  =  q(1-P_{eq}) \cdot \sum_{i=0}^{m}b_{i,0} + q \cdot \frac{(1-q)(1-P_{eq})}{q} \cdot \sum_{i=0}^{m}b_{i,0}& \\
            =  (1-P_{eq})[q + (1-q)] \cdot \sum_{i=0}^{m}b_{i,0} &\\
            =  (1-P_{eq}) \cdot \sum_{i=0}^{m}b_{i,0} &
\end{array}
\end{equation}
Employing the normalization condition, after some mathematical
manipulations, and remembering the relation $
\sum_{i=0}^{m}b_{i,0} = \frac{b_{0,0}}{1-P_{eq}} $, it is possible
to obtain:
\begin{equation}\small
\begin{array}{lcl}
1 & = & \sum_{i=0}^{m}\sum_{k=0}^{W_i-1} b_{i,k} + b_I \\
  & = & \frac{b_{0,0}}{2} \left[ W\left( \sum_{i=0}^{m-1}(2P_{eq})^i + \frac{(2P_{eq})^m}{1-P_{eq}} \right) +
  \frac{1}{1-P_{eq}}\right]+\\

    &&    + \frac{(1-q)(1-P_{eq})}{q} \cdot \sum_{i=0}^{m}b_{i,0} \\
  & = & \frac{b_{0,0}}{2} \left[ W\left( \sum_{i=0}^{m-1}(2P_{eq})^i + \frac{(2P_{eq})^m}{1-P_{eq}} \right) +
  \frac{1}{1-P_{eq}}\right]+\\
    &&    + \frac{(1-q)(1-P_{eq})}{q} \cdot \frac{b_{0,0}}{1-P_{eq}} \\
  & = & \frac{b_{0,0}}{2} \left[ W\left( \sum_{i=0}^{m-1}(2P_{eq})^i + \frac{(2P_{eq})^m}{1-P_{eq}} \right) + \frac{1}{1-P_{eq}}
        + \frac{2(1-q)}{q} \right] \\
\end{array}
\end{equation}
Normalization condition yields the following equation for
computation of $b_{0,0}$:
\begin{equation}
\label{eqb_00_norm}
\begin{array}{ll}
  b_{0,0}=  \frac{2}{\frac{W[(1-P_{eq})\sum_{i=0}^{m-1}(2P_{eq})^i + (2P_{eq})^m] + 1 + 2\frac{1-q}{q}(1-P_{eq})}{1-P_{eq}}} & \\
         =  \frac{2(1-P_{eq})}
                   {(W-P_{eq}W)\frac{1-(2P_{eq})^m}{(1-2P_{eq})} + W(2P_{eq})^m + 1 + 2\frac{(1-q)(1-P_{eq})}{q}} &  \\
         =  \frac{2(1-P_{eq})(1-2P_{eq})q}
                   {q[(W+1)(1-2P_{eq}) + WP_{eq}(1-(2P_{eq})^m)] +
                   2(1-q)(1-P_{eq})(1-2P_{eq})}&
\end{array}
\end{equation}
As a side note, if $q\rightarrow 1$, i.e., the
station is approaching saturated traffic conditions and assuming packet
transmission errors are only due to collisions, i.e., $P_e=0$ and
$P_{eq}= P_{col}$, (\ref{eqb_00_norm}) we get:
\begin{equation}
\label{eq:b_00_norm_sat}
\begin{array}{ll}
\lim_{q\rightarrow 1} b_{0,0} \rightarrow
\frac{2(1-P_{col})(1-2P_{col})}{(W+1)(1-2P_{col}) +
WP_{col}(1-(2P_{col})^m)}
\end{array}
\end{equation}
which corresponds to the stationary state probability $b_{0,0}$ found by
Bianchi \cite{Bianchi} under saturated conditions.

Equ.~(\ref{eqb_00_norm}) is then used to compute $\tau$, the
probability that a station starts a transmission in a randomly
chosen time slot. In fact, taking into account that a packet
transmission occurs when the backoff counter reaches zero, we
have:
\begin{equation}\label{eq.tau}
\begin{array}{ll}
  \tau & =\sum_{i=0}^{m}b_{i,0} =\frac{b_{0,0}}{1-P_{eq}}= \\
      = & \frac{2(1-2P_{eq})q}{q[(W+1)(1-2P_{eq}) + WP_{eq}(1-(2P_{eq})^m)] + 2(1-q)(1-P_{eq})(1-2P_{eq})} \\
\end{array}
\end{equation}
Once again as a consistency check, note that if $m=0$, i.e., no exponential
backoff is employed, $q\rightarrow 1$, i.e., the station is
approaching saturated traffic conditions, and under the condition that packet transmission
errors are only due to collisions, i.e., $P_e=0$ and $P_{eq}=
P_{col}$, (\ref{eq.tau}) we get:
\begin{equation}
\label{eq:tau_norm_sat}
\lim_{q\rightarrow 1} \tau \rightarrow
\frac{2}{W+1}
\end{equation}
which is the result found in~\cite{Ho} for the constant backoff
setting, showing that the transmission probability $\tau$ is
independent of the collision probability $P_{col}$.

The collision probability needed to compute $\tau$ can be found
considering that using a 2-way hand-shaking mechanism, a packet from a
transmitting station encounters a collision if in a given time slot, at
least one of the remaining $(N-1)$ stations transmits
simultaneously another packet, and there is no capture. In our
model, we assume that capture is a subset of the collision events.
This is indeed justified by the fact that there is no capture
without collision, and that capture occurs only because of
collisions between a certain number of transmitting stations
attempting to transmit simultaneously on the channel.
\begin{equation}\label{eq.col}
P_{col} = 1-(1-\tau)^{N-1}-P_{cap}
\end{equation}
As far as the capture effects are concerned, we resort to the
mathematical formulation proposed in \cite{zorzi_rao,Spasenovski}.

We consider a scenario in which $N$ stations are uniformly
distributed in a circular area of radius $R$, and transmit toward
a common access point placed in the center of such an area. When signal
transmission is affected by Rayleigh fading, the instantaneous
power of the signal received by the receiver placed at a mutual
distance $r_i$ from the transmitter is exponentially distributed
as:
\[
f(x)=\frac{1}{p_o}e^{-\frac{x}{p_o}}, x>0
\]
whereby $p_o$ is the local-mean power determined by the
equation:
\begin{equation}\label{eq.transmission_equat}
p_o=A\cdot r_i^{-n_p} P_{tx}
\end{equation}
where, $n_p$ is the path-loss exponent (which
is typically greater or equal to $3.5$ in indoor propagation
conditions in the absence of the direct signal path), $P_{tx}$ is
the transmitted power, and $A\cdot r_i^{-n_p}$ is the
deterministic path-loss \cite{Rappaport}. Both $A$ and $P_{tx}$
are identical for all transmitted frames.

Under the hypothesis of power-controlled stations in
infrastructure mode and Rayleigh fading, the capture probability
conditioned on $i$ interfering frames can be defined as follows:
\begin{equation}\label{eq.capture_conditional}
P_{cp}\left(\gamma>z_o
g(S_f)|i\right)=\frac{1}{{\left[1+z_{0}g(S_{f})\right]}^{i}}
\end{equation}
whereby, $\gamma$, defined as
\begin{equation}\label{eq.defgamma}
\gamma=P_u/\sum_{k=1}^{i}P_k
\end{equation}
is the ratio of the power $P_u$ of the useful signal and the sum
of the powers of the $i$ interfering channel contenders
transmitting simultaneously $i$ frames, $g(S_f)$ is the inverse of
the processing gain of the correlation receiver, and $z_0$ is the
capture ratio, i.e., the value of the signal-to-interference power
ratio identifying the capture threshold at the receiver. Notice
that (\ref{eq.capture_conditional}) signifies the fact that
capture probability corresponds to the probability that the power
ratio $\gamma$ is above the capture threshold $z_{0}g(S_{f})$
which considers the inverse of the processing gain $g(S_{f})$. For
Direct Sequence Spread Spectrum (DSSS) using a 11-chip spreading
factor ($S_f=11$), we have $g(S_{f})=\frac{2}{3S_f}$
\cite{Pursley}.

Upon defining the probability of generating exactly $i+1$
interfering frames over $N$ contending stations in a generic slot
time:
\[
{N \choose i+1}\tau^{i+1}(1-\tau)^{N-i-1}
\]
the frame capture probability $P_{cap}$ can be obtained as follows:
\begin{equation}\label{capture_probability}
P_{cap}=\sum_{i=1}^{N-1}{N \choose
i+1}\tau^{i+1}(1-\tau)^{N-i-1}P_{cp}\left(\gamma>z_o
g(S_f)|i\right)
\end{equation}
Putting together
Equ.s~(\ref{eq.equ}),~(\ref{eq.tau}),~(\ref{eq.col}),
and~(\ref{capture_probability}) the following nonlinear system can
be defined and solved numerically, obtaining the values of $\tau$,
$P_{col}$, $P_{cap}$, and $P_{eq}$:
\begin{equation}\label{eq.system}
\left\{ \begin{array}{ll}
\tau  =\frac{2(1-2P_{eq})q}{q[(W+1)(1-2P_{eq}) + WP_{eq}(1-(2P_{eq})^m)] + 2(1-q)(1-P_{eq})(1-2P_{eq})}\\
P_{col} = 1-(1-\tau)^{N-1} -P_{cap}\\
P_{eq}   =
P_{col}+P_e-P_e\cdot  P_{col}\\
P_{cap}=\sum_{i=1}^{N-1}{N \choose
i+1}\tau^{i+1}(1-\tau)^{N-i-1}\frac{1}{{(1+z_{0}g(S_{f}))}^{i}}
\end{array} \right.
\end{equation}
The final step in the analysis is the computation of the
normalized system throughput, defined as the fraction of time the
channel is used to successfully transmit payload bits:
\begin{equation}\small
\label{eq.system2} S = \frac{P_t \cdot P_s\cdot
(1-P_e)E[PL]}{(1-P_t
)\sigma+P_t(1-P_s)T_c+P_tP_s(1-P_e)T_s+P_tP_sP_eT_e}
\end{equation}
where,
\begin{itemize}
\item $P_t$ is the probability that there is at least one
transmission in the considered time slot, with $N$ stations
contending for the channel, each transmitting with probability
$\tau$:
\begin{equation}
\label{equat_Pt} P_t=1-(1-\tau)^N
\end{equation}
\item $P_s$ is the conditional probability that a packet
transmission occurring on the channel is successful. This event
corresponds to the case in which exactly one station transmits in
a given time slot, or two or more stations transmit simultaneously
and capture by the desired station occurs. These conditions yields
the following probability:
\begin{equation}
\label{equat_PS}
P_s=\frac{N\tau(1-\tau)^{N-1}+P_{cap}}{P_t}
\end{equation}
\item $T_c$, $T_e$ and $T_s$ are the average times a channel is
sensed busy due to a collision, error affected data frame
transmission time and successful data frame transmission times,
respectively. Knowing the time durations for ACK frames, ACK
timeout, DIFS, SIFS, $\sigma$, data packet length ($PL$) and PHY
and MAC headers duration ($H$), and propagation delay $\tau_p$,
$T_c$, $T_s$, and $T_e$ can be computed as follows \cite{kong}:
\begin{equation}
 \begin{array}{ll}
T_c       &= H+PL+ACK_{timeout}              \\
T_e       &= H+PL+ACK_{timeout} \\
T_s       &= H+PL+SIFS+ \tau_p+ACK+DIFS +\tau_p\\
\end{array}
\end{equation}
\item $E[PL]$ is the average packet payload length.

\item $\sigma$ is the duration of an empty time slot.
\end{itemize}

The setup described above is used in Section
\ref{Simulation_results_section} for DCF simulation at the MAC
layer.
\subsection{Modelling offered load and estimation of probability $q$}
In our analysis, the offered load related to each station is
characterized by parameter $\lambda$ representing the rate at
which packets arrive at the station buffer from the upper layers,
and measured in $pkt/s$. The time between two packet arrivals is
defined as \textit{interarrival time}, and its mean value is
evaluated as $\frac{1}{\lambda}$. One of the most commonly used traffic
models assumes packet arrival process is Poisson. The resulting interarrival
times are exponentially distributed.

In the proposed model, we need a probability $q$ that indicates if
there is at least one packet to be transmitted in the queue.
Probability $q$ can be well approximated in a situation with small
buffer size \cite{hamilton} through the following relation:
\begin{equation}
\label{eq:q_prob} q = 1 - e^{- \lambda E[S_{ts}]}
\end{equation}
where, $E[S_{ts}]$ is the \textit{expected time per slot}, which is
useful to relate the state of the Markov chain with the actual
time spent in each state.

A more accurate model can be derived upon considering different
values of $q$ for each backoff state. However, a
reasonable solution consists in using a mean probability valid for
the whole Markov model~\cite{hamilton}, derived from $E[S_{ts}]$.
The value of $E[S_{ts}]$ can be obtained by resorting to the
durations and the respective probabilities of the idle slot
($\sigma$), the successful transmission slot ($T_s$), the error
slot due to collision ($T_c$), and the error slot due to channel
($T_e$), as follows:
\begin{equation}
\label{equ_E_Sts}
\begin{array}{rcl}
E[S_{ts}] & = & (1-P_t) \cdot \sigma + P_t(1-P_s) \cdot T_c +\\
     &   &  +P_t P_s P_e \cdot T_e + P_t P_s (1-P_e) \cdot T_s
\end{array}
\end{equation}
Upon recalling that packet inter-arrival times are exponentially distributed,
we can use the average slot time to calculate the probability $q$
that in such a time interval a given station receives a packet
from upper layers in its transmission queue. The probability that
in a generic time $T$, $k$ events occur, is:
\begin{equation}
P\{a(T) = k\} = e^{- \lambda T} \frac{(\lambda T)^k}{k!}
\end{equation}
from which we obtain the relation (\ref{eq:q_prob}) referred to earlier:
\begin{equation}
\label{eq:q_prob2} q = 1 - P\{a(E[S_{ts}]) = 0\} = 1 - e^{-
\lambda E[S_{ts}]}
\end{equation}
\section{Simulation Results and Model Validations}
\label{Simulation_results_section}
\begin{table}\caption{Typical network parameters}
\begin{center}
\begin{tabular}{|c|c|}\hline
\hline MAC header & 24 bytes\\
\hline PHY header & 16 bytes\\
\hline Payload size & 1024 bytes\\
\hline ACK & 14 bytes\\
\hline RTS & 20 bytes\\
\hline CTS & 14 bytes\\
\hline
\hline $\tau_p$ & 1 $\mu s$\\
\hline Slot time & 20 $\mu s$\\
\hline SIFS & 10 $\mu s$\\
\hline DIFS & 50 $\mu s$\\
\hline EIFS & 300 $\mu s$\\

\hline ACK timeout & 300 $\mu s$\\
\hline CTS timeout & 300 $\mu s$\\
\hline\hline
\end{tabular}
 \label{tab.design.times}
\end{center}
\end{table}
This section focuses on simulation results for validating the
theoretical models and derivations presented in the previous
sections. We have developed a C++ simulator modelling both the DCF
protocol details in 802.11b and the backoff procedures of a
specific number of independent transmitting stations. The
simulator is designed to implement the main tasks accomplished at
both MAC and PHY layers of a wireless network in a more versatile
and customizable manner than ns-2, where the lack of a complete
physical layer makes difficult a precise configuration at this
level.

The simulator considers an Infrastructure BSS (Basic Service Set)
with an AP and a certain number of mobile stations which
communicates only with the AP. For the sake of simplicity, inside
each station there are only three fundamental working levels:
traffic model generator, MAC and PHY layers. Traffic is generated
following the exponential distribution for the packet interarrival
times. Moreover, the MAC layer is managed by a state machine which
follows the main directives specified in the
standard~\cite{standard_DCF_MAC}, namely waiting times (DIFS,
SIFS, EIFS), post-backoff, backoff, basic and RTS/CTS access mode.

Typical MAC layer parameters for IEEE802.11b are given in
Table~\ref{tab.design.times}~\cite{standard_DCF_MAC}. In so far as
the computation of the FER is concerned, it should be noted that
data transmission rate of various packet types differ. For
simplicity, we assume that data packets transmitted by different
stations are affected by the same FER. This way, channel errors on
the transmitted packets can be accounted for as it is done within
ns-2~\cite{xiuchao}. In other words, a uniformly distributed
binary random variable $X_e$ is generated in order to decide if a
transmitted packet is received erroneously. The statistic of such
a random variable is $P(X_e=1)=P_e(SNR)$ (as specified in
(\ref{fer_1})), and $P(X_e=0)=1-P_e(SNR)$.

As far as capture effects are concerned, the investigated scenario
is as follows. In our simulator, $N$ stations are randomly placed
in a circular area of radius $R$ (in the simulation results
presented below we assume $R=10$m), while the AP is placed at the
center of the circle. When two or more station transmissions collide, the
value of $\gamma$ as defined in (\ref{eq.defgamma}) is evaluated
for any transmitting station given their relative distance $r_i$
from the AP. Let $\gamma_j$ be the value of $\gamma$ for the
$j$-th transmitting station among the $i+1$ colliding stations.
The power between a transmitter and a receiver is evaluated in
accordance to (\ref{eq.transmission_equat}) with a path-loss
exponent equal to $3.5$. The values of $\gamma_j$ for each
colliding station are compared with the threshold $z_o\cdot
g(S_f)$. Then, the transmitting station for which $\gamma_j$ is
above the threshold captures the channel.

The physical layer (PHY) of the basic 802.11b standard is based on
the spread spectrum technology. Two options are specified, the
Frequency Hopped Spread Spectrum (FHSS) and the Direct
Sequence Spread Spectrum (DSSS). The FHSS uses Frequency Shift Keying
(FSK) while the DSSS uses Differential Phase Shift Keying (DPSK)
or Complementary Code Keying (CCK). The 802.11b employs DSSS at
various rates including one employing CCK encoding 4 and 8 bits on
one CCK symbol. The four supported data rates in 802.11b are 1, 2,
5.5 and 11 Mbps.

In particular, the following data are transmitted at lowest rate
(1 Mbps) in IEEE802.11b:
PLCP=16 bytes (PLCP plus header), ACK Headers=16 bytes, ACK=14
bytes, RTS=20 bytes, CTS=14 bytes (RTS and CTS are only for four way handshake).

The FER as a function of the SNR can be computed as follows:
\begin{equation}\label{fer_1}\small
P_e(SNR)=1-\left[1-P_e(PLCP,SNR)\right]\cdot
\left[1-P_e(DATA,SNR)\right]
\end{equation}
where,
\begin{equation}\label{fer_2}\small
P_e(PLCP,SNR)=1-\left[1-P_b(BPSK,SNR)\right]^{8\times PLCP},
\end{equation}
and
\begin{equation}\label{fer_3}\small
P_e(DATA,SNR)=1-\left[1-P_b(TYPE,SNR)\right]^{8\times (DATA+MAC)}.
\end{equation}
$P_b(BPSK,SNR)$ is the BER as a function of SNR for the lowest
data transmit rate employing DBPSK modulation, DATA denotes the
data block size in bytes, and any other constant byte size in
above expression represents overhead. Note that the FER,
$P_e(SNR)$, implicitly depends on the modulation format used.
Hence, for each supported rate, one curve for $P_e(SNR)$ as a
function of SNR can be generated. $P_b(TYPE,SNR)$ is modulation
dependent whereby the parameter $TYPE$ can be any of the following
$TYPE\in \{DBPSK,DQPSK,CCK5.5,CCK11\}$\footnote{The acronyms are
short for Differential Binary Phase Shift Keying, Differential
Quadrature Phase Shift Keying and Complementary Code Keying,
respectively.}.

For DBPSK and DQPSK modulation formats, $P_b(TYPE,SNR)$ can be
well approximated by \cite{SimonAlouini}:
\figura{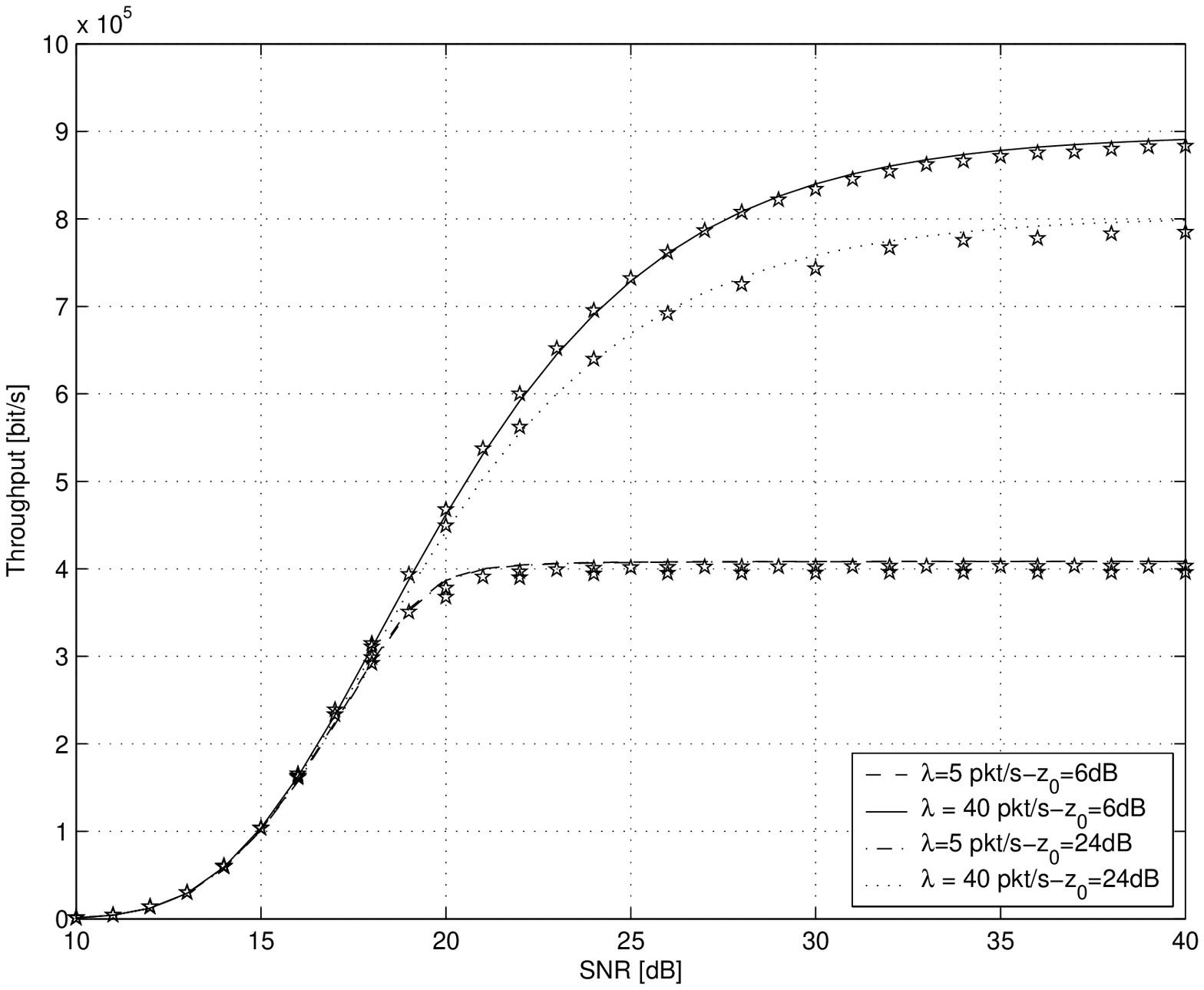}{Theoretical and
simulated throughput for the 2-way mechanism as a function of the
SNR, for two different packet rates and capture thresholds, shown
in the legend. Simulated points are identified by star-markers
over the respective theoretical curves. Payload size is 1024
bytes, while the number of contending stations is
$N=10$.}{Throgh_SNR_lambda_par_N10}
\begin{equation}\small
\label{ber_function}
\frac{2}{\max(\log_{2}M,2)}\sum_{i=1}^{\max(\frac{M}{4},1)}\frac{1}{\pi}\int_{0}^{\frac{\pi}{2}}
\frac{1}{1+\frac{\gamma\log_{2}M}{\sin^2\theta}\sin^2\left(\frac{(2i-1)\pi}{M}\right)}d\theta
\end{equation}
whereby, $M$ is the number of bits per modulated symbols, $\gamma$
is the signal-to-noise ratio, and $\theta$ is the signal direction
over the Rayleigh fading channel.

We note
that the proposed DCF model is valid with any other PHY setup.
Actually, all we need is simply the packet error rate probability
$P_e(DATA,SNR)$ in (\ref{fer_1}) for the specific PHY and channel
transmission conditions model. All the mathematical derivations
proposed in this paper are specified with respect to
$P_e(DATA,SNR)$, and so they are valid for any kind of
transmission model at the PHY layer.

In what follows, we shall present theoretical and simulation
results for the lowest supported data rate. We note that by
repeating the process, similar curves can be generated for all
other types of modulation formats. All we need is really the BER
as a function of SNR for each modulation format and the
corresponding raw data rate over the channel. If the terminals use
rate adaptation, then under optimal operating condition, the
achievable throughput for a given SNR is the maximum over the set
of modulation formats supported. We have verified a close match
between theoretical and simulated performances for other
transmitting data rates as well. In the results presented below we
assume the following values for the contention window:
$CW_{min}=32$, $m=5$, and $CW_{max} = 2^m\cdot CW_{min} = 1024$.

Fig.~\ref{Throgh_SNR_lambda_par_N10} shows the behavior of the
throughput for the 2-way mechanism as a function of the SNR, for
two different capture thresholds, $z_0$, and packet rates,
$\lambda$, as noted in the legend, and for $N=10$ transmitting
stations. Simulated points are marked by star on the respective
theoretical curves. Throughput increases
as a function of the three parameters SNR, $\lambda$, and $z_0$ until
saturation point.
\figura{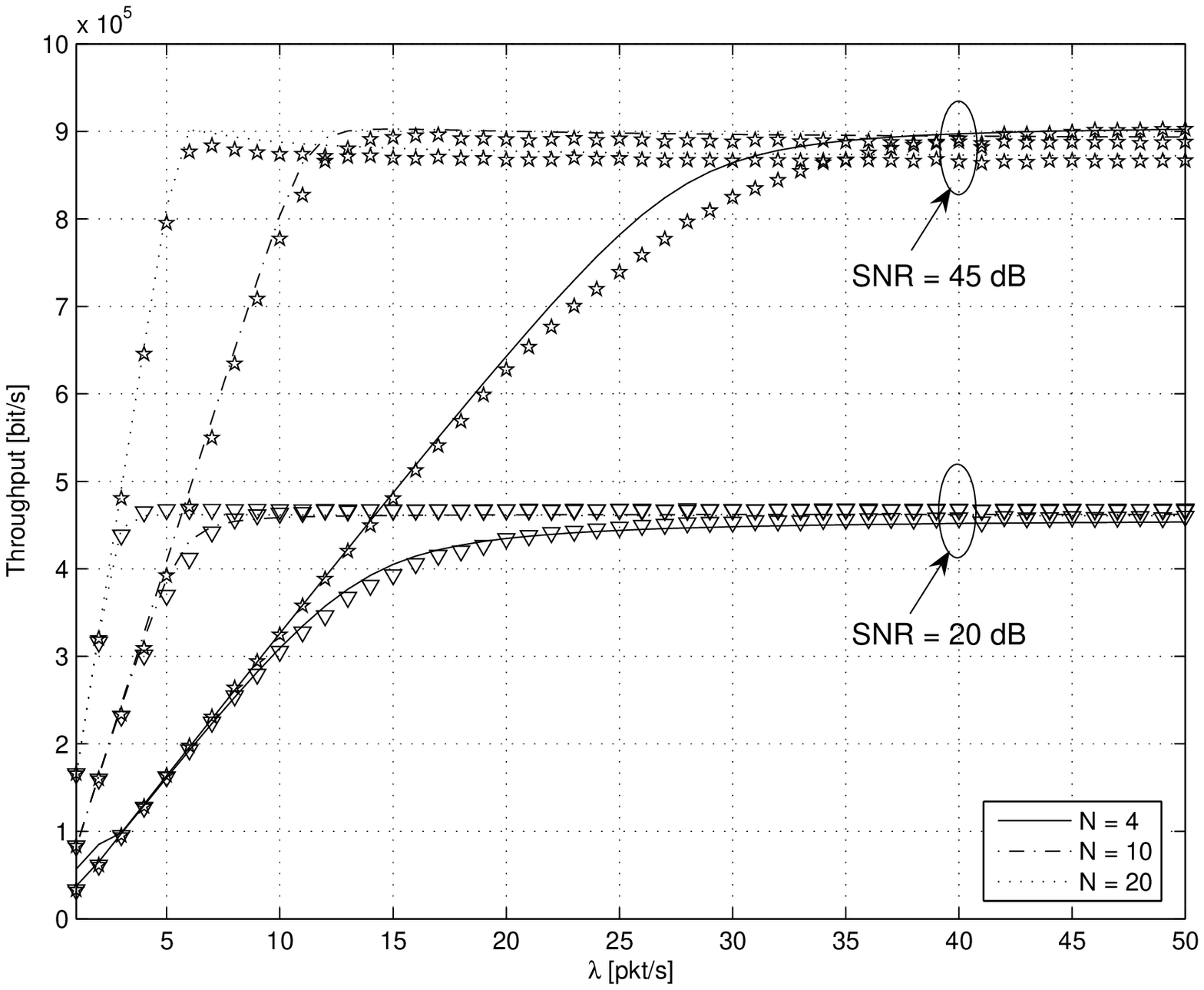}{Theoretical and
simulated throughput for the 2-way mechanism as a function of the
packet rate $\lambda$, for three different number of contending
stations and two different values of SNR as noted in the legend.
Capture thresholds is $z_0=6$dB. Simulated points are identified
by star-markers over the respective theoretical curves. Payload
size is 1024 bytes.}{Throgh_lambda_par_z0_6dB}

Fig.~\ref{Throgh_lambda_par_z0_6dB} shows the behavior of the
throughput as a function of $\lambda$, i.e., the packet rate, for
three different values of the number of contending stations and
for two values of SNR. The capture threshold is $z_0=6$dB. Beside
noting the throughput improvement achievable for high SNR, notice
that for a specified number of contending stations, the throughput
manifests a linear behavior for low values of packet rates with a
slope depending mainly on the number of stations $N$. However, for
increasing values of $\lambda$, the saturation behavior occurs
quite fast. Notice that, as exemplified in~(\ref{eq:q_prob2}),
$q\rightarrow 1$ as $\lambda\rightarrow\infty$. Actually,
saturated traffic conditions are achieved quite fast for values of
$\lambda$ on the order of ten packets per second with a number of
contending stations greater than or equal to 10.

Fig.~\ref{Throgh_lambda_par_z0_24dB} shows the behavior of the
throughput as a function of $\lambda$, for three different values
of the number of contending stations and for two different SNRs.
The capture threshold is $z_0=24$dB. We can draw conclusions
similar to the ones derived for
Fig.~\ref{Throgh_lambda_par_z0_6dB}. Upon comparing the curves
shown in Figs~\ref{Throgh_lambda_par_z0_6dB}
and~\ref{Throgh_lambda_par_z0_24dB}, it is easily seen that
capture effects allow the system throughput to be almost the same
independently from the number of stations in saturated conditions,
i.e., for high values of $\lambda$.

Fig.~\ref{Throgh_lambda_par_z0_24dB} also shows the presence of a
peak in the throughput as a function of $\lambda$, which
characterizes the transition between the linear and saturated
throughput. Such a peak tends to manifest itself for increasing
values of $\lambda$ as the number of stations $N$ increases. A
comparative analysis of the curves shown in
Figs~\ref{Throgh_lambda_par_z0_6dB}
and~\ref{Throgh_lambda_par_z0_24dB} reveals that the peak of the
throughput tends to disappear because of the presence of capture
effects during transmission.

Notice also that the throughput behavior exhibits saturation quite fast
for low values of SNR independently from
capture effects; this is essentially due to the fact that channel
propagation errors tend to dominate over both collisions and
capture when the SNR is low.
\figura{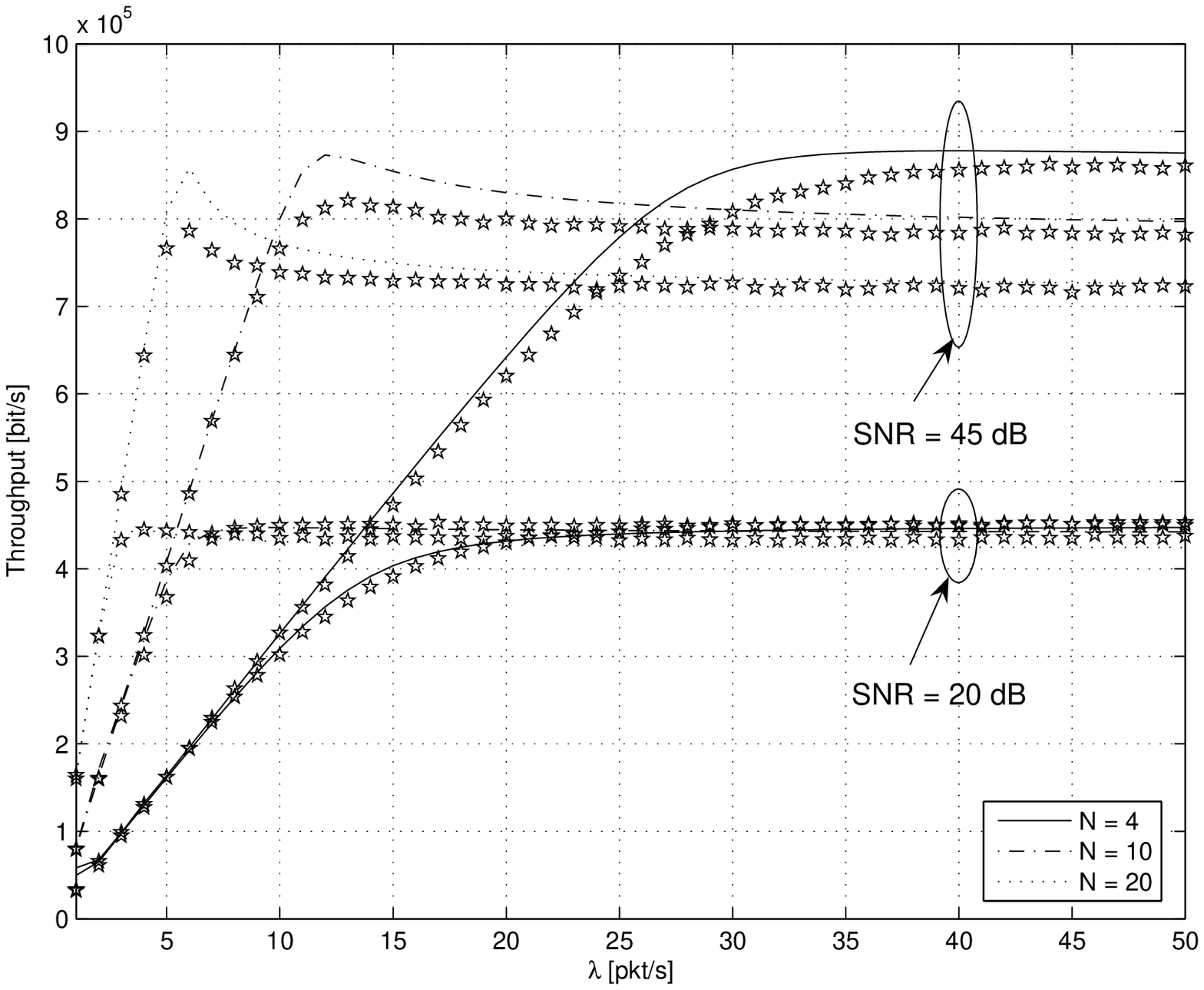}{Theoretical and
simulated throughput for the 2-way mechanism as a function of the
packet rate $\lambda$, for three different number of contending
stations and two different values of SNR as noted in the legend.
Capture threshold is $z_0=24$dB. Simulated points are identified
by star-markers over the respective theoretical curves. Payload
size is 1024 bytes.}{Throgh_lambda_par_z0_24dB}

In order to assess throughput performances as a function of the
payload size, Fig.~\ref{Throgh_lambda_par_variN_z0} shows the
behavior of the throughput as a function of $\lambda$, for a
payload size equal to 128 bytes. The other simulation parameters
are noted in the legend. Beside noting similar conclusions on the
effects of both capture and number of contending stations over the
throughput behavior as for the previous mentioned figures, notice
also that a smaller payload size reduces proportionally the slope
of the throughput before saturated conditions, and the maximum
achievable throughput in saturated conditions. This issue will be
dealt with in the next section.
\figura{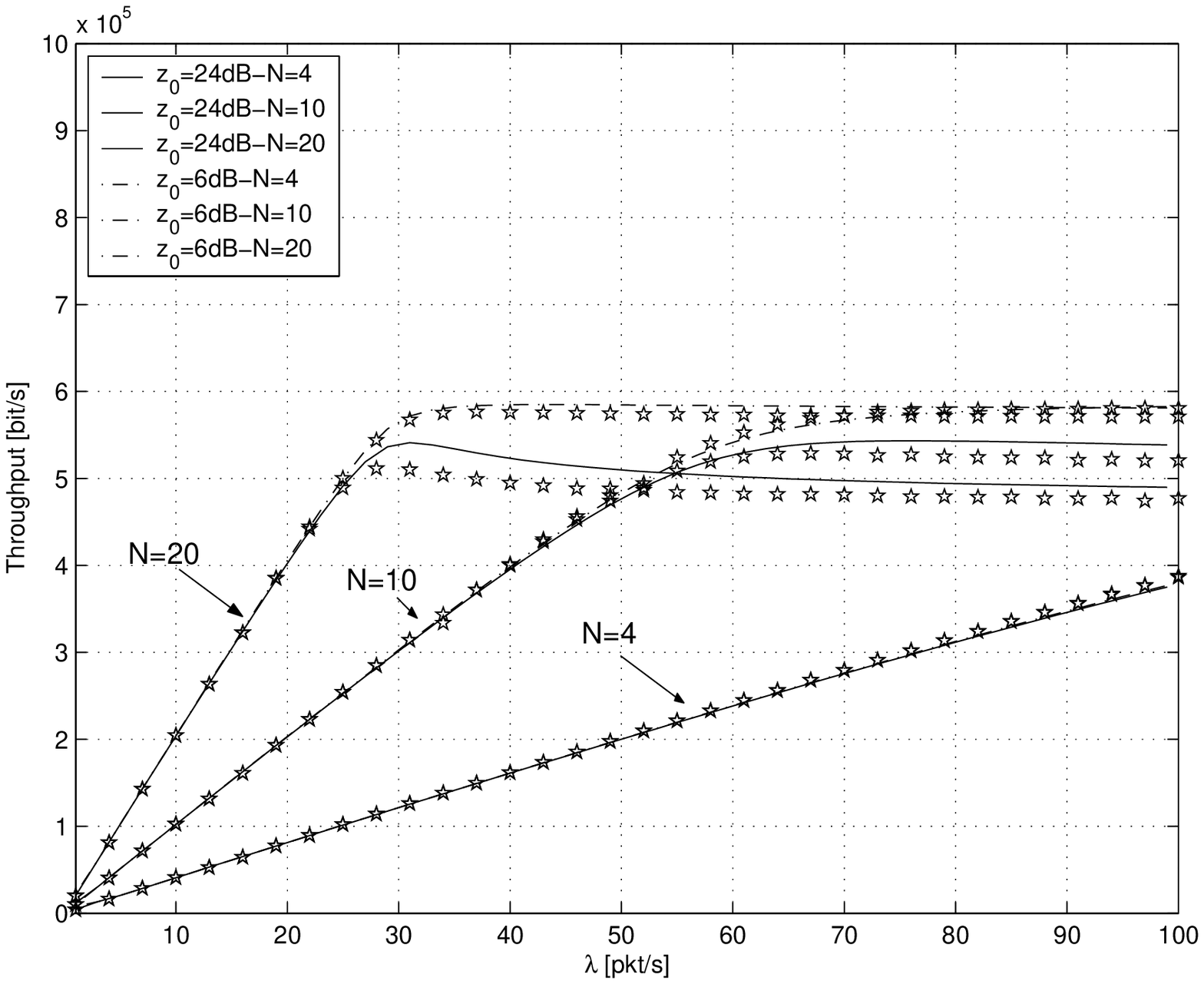}{Theoretical and
simulated throughput for the 2-way mechanism as a function of the
packet rate $\lambda$, for three different number of contending
stations and two different capture thresholds noted in the legend.
Simulated points are identified by star-markers over the
respective theoretical curves. Payload size is 128 bytes, while
SNR=$45$dB.}{Throgh_lambda_par_variN_z0}
\subsection{Simulation results within ns-2}
For the sake of validating the proposed model and our developed simulator against a widely
adopted network simulator, while providing simulation results for
a new set of parameters, we have performed new simulations using ns-2 (version 2.29)
for throughput evaluations. Due
to the lack of a complete PHY layer implementation, we made some
modifications in the original ns-2 source code in order to account
for a specific FER on the transmitted packets. In this respect, we
adopted the suggestions proposed in \cite{xiuchao} for simulating
a 802.11b channel within ns-2. Briefly, channel errors on the
transmitted packets are accounted for by using a uniformly
distributed binary random variable $X_e$ with the following
probabilities: $P(X_e=1)=P_e(SNR)$ (as specified in
(\ref{fer_1})), and $P(X_e=0)=1-P_e(SNR)$. For simplicity, we
assume that data packets transmitted by different stations are
affected by the same FER.

The employed propagation model is identified by the label
\verb"Propagation/Shadowing" within ns-2. We considered a
path-loss exponent equal to 3.5 with a zero standard deviation in
compliance with the theoretical capture scenario described above.
Notice that the standard ns-2 simulator provides only Ad-Hoc
support and related routing protocols; for this reason we applied
a patch for implementing an Infrastructure BSS without any routing
protocol between the various involved stations. This patch is
identified by the acronym \textit{NOAH}, standing for NO Ad-Hoc,
and can be downloaded from
\textrm{http://icapeople.epfl.ch/widmer/uwb/ns-2/noah/}.

In connection with the employed traffic model, we used an
exponential distribution for simulating packet interarrival times.

Throughput evaluation is accomplished by averaging over $100$
sample scenarios, whereby any transmitting scenario considers a
set of $N$ randomly distributed (with a uniform pdf) stations over a circular
area of radius 10 m. Simulation results along with theoretical
curves are shown in Figs~\ref{N10_ideal}-\ref{N5_10_capture}. The
employed parameters are noted in the respective legends of each
figure. While the main conclusions related to the throughput
behavior as a function of the number of contending stations,
capture threshold, traffic load, and FER, are similar to the ones
already derived above with our C++ simulator, here we notice a
good agreement between theoretical and ns-2 simulation results for
a wide range of packet arrival rates, $\lambda$.

Figs~\ref{N10_ideal}-\ref{N5_ideal} also depict the theoretical
saturated throughput derived by Bianchi in~\cite{Bianchi} in order
to underline a good matching between our theoretical model, valid
for any traffic load in the investigated range of $\lambda$ up to
the saturation condition for $\lambda\rightarrow\infty$.
\figura{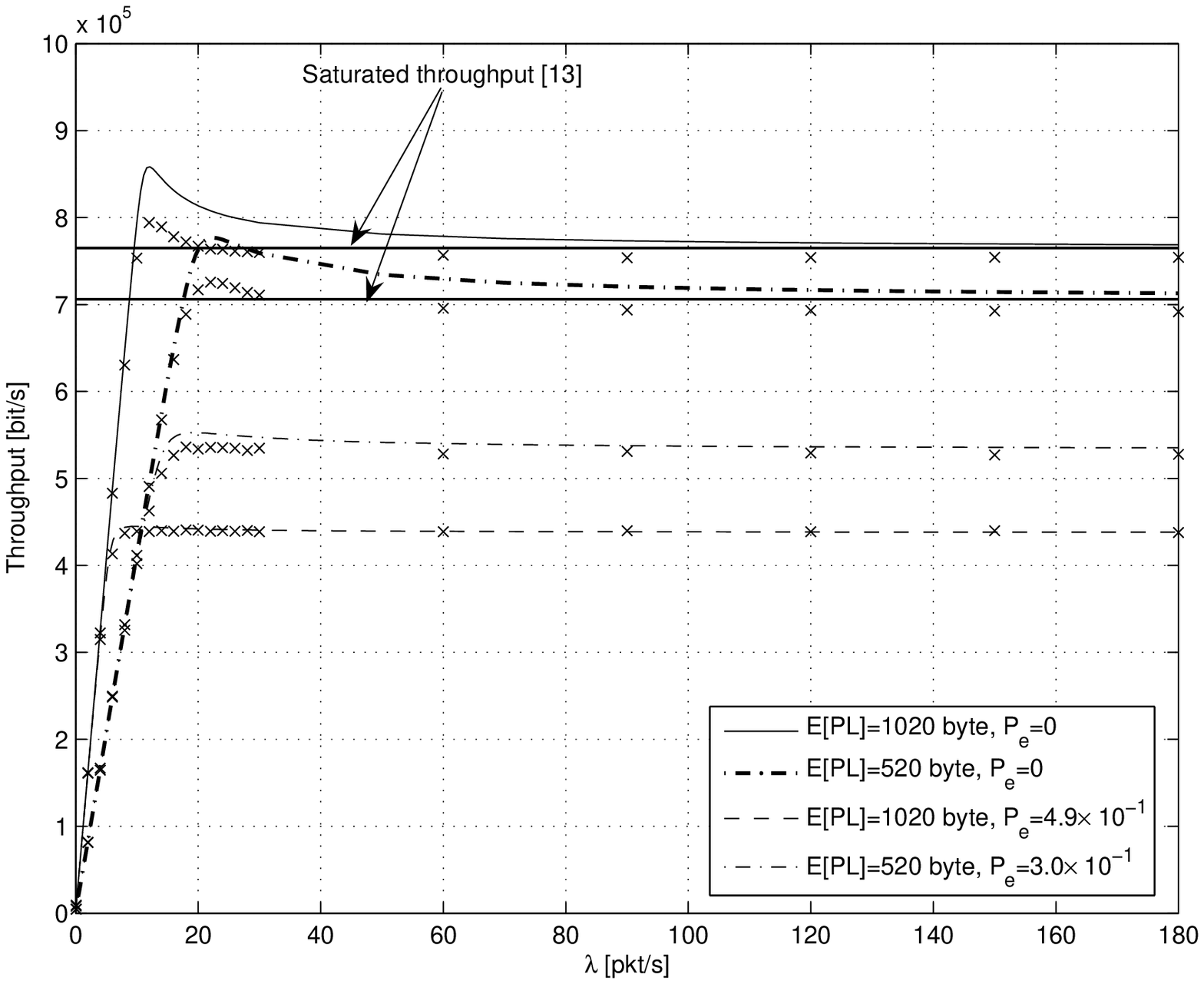}{Theoretical and
simulated (ns-2) throughput for the 2-way mechanism as a function
of the packet rate $\lambda$, for two different payload sizes and
packet error probabilities, shown in the legend, and without
capture. Simulated points are identified by cross-markers over the
respective theoretical curves. The number of contending stations
is $N=10$.}{N10_ideal}
\figura{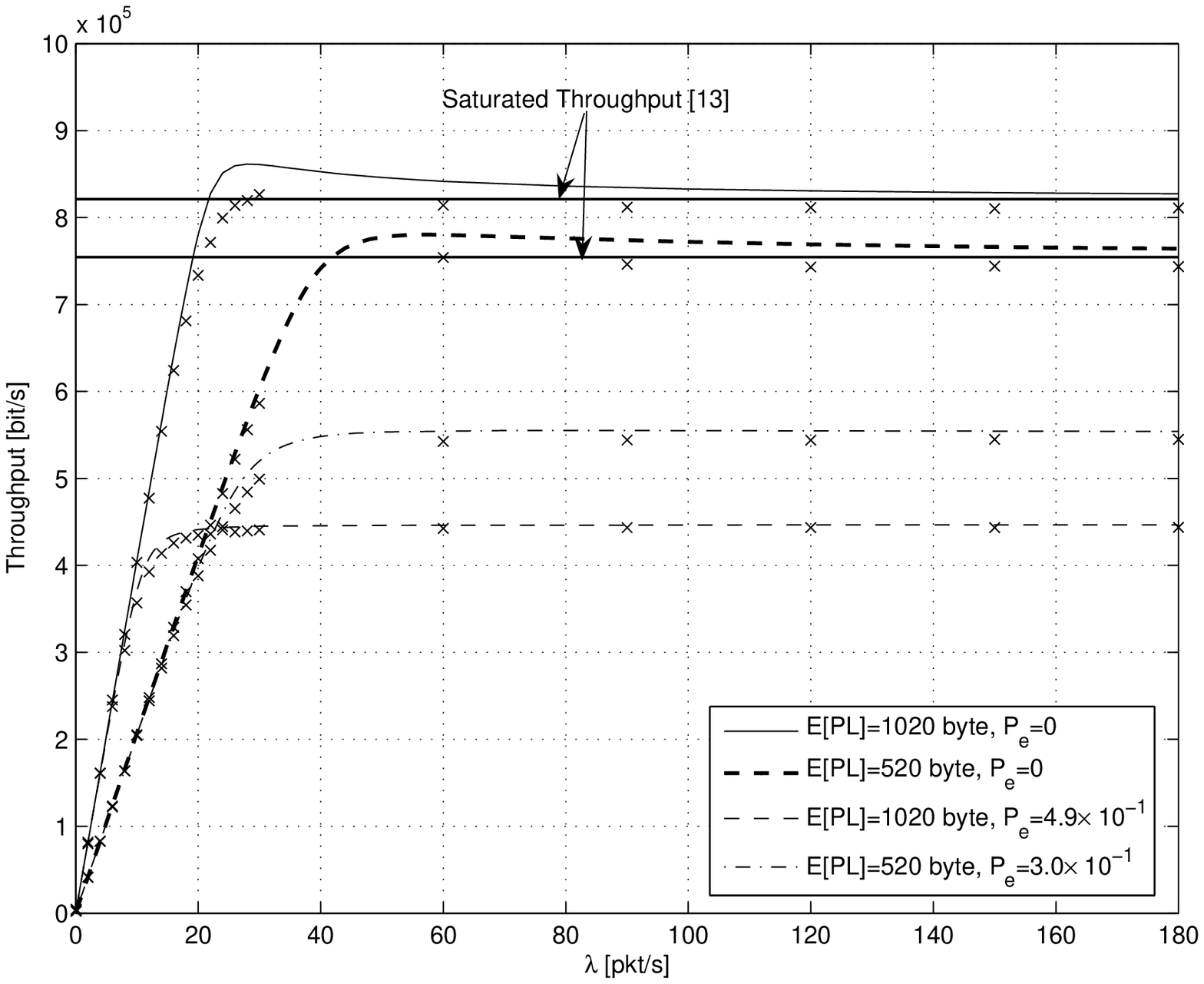}{Theoretical and
simulated (ns-2) throughput for the 2-way mechanism as a function
of the packet rate $\lambda$, for two different payload sizes and
packet error probabilities, shown in the legend, and without
capture. Simulated points are identified by cross-markers over the
respective theoretical curves. The number of contending stations
is $N=5$.}{N5_ideal}
\figura{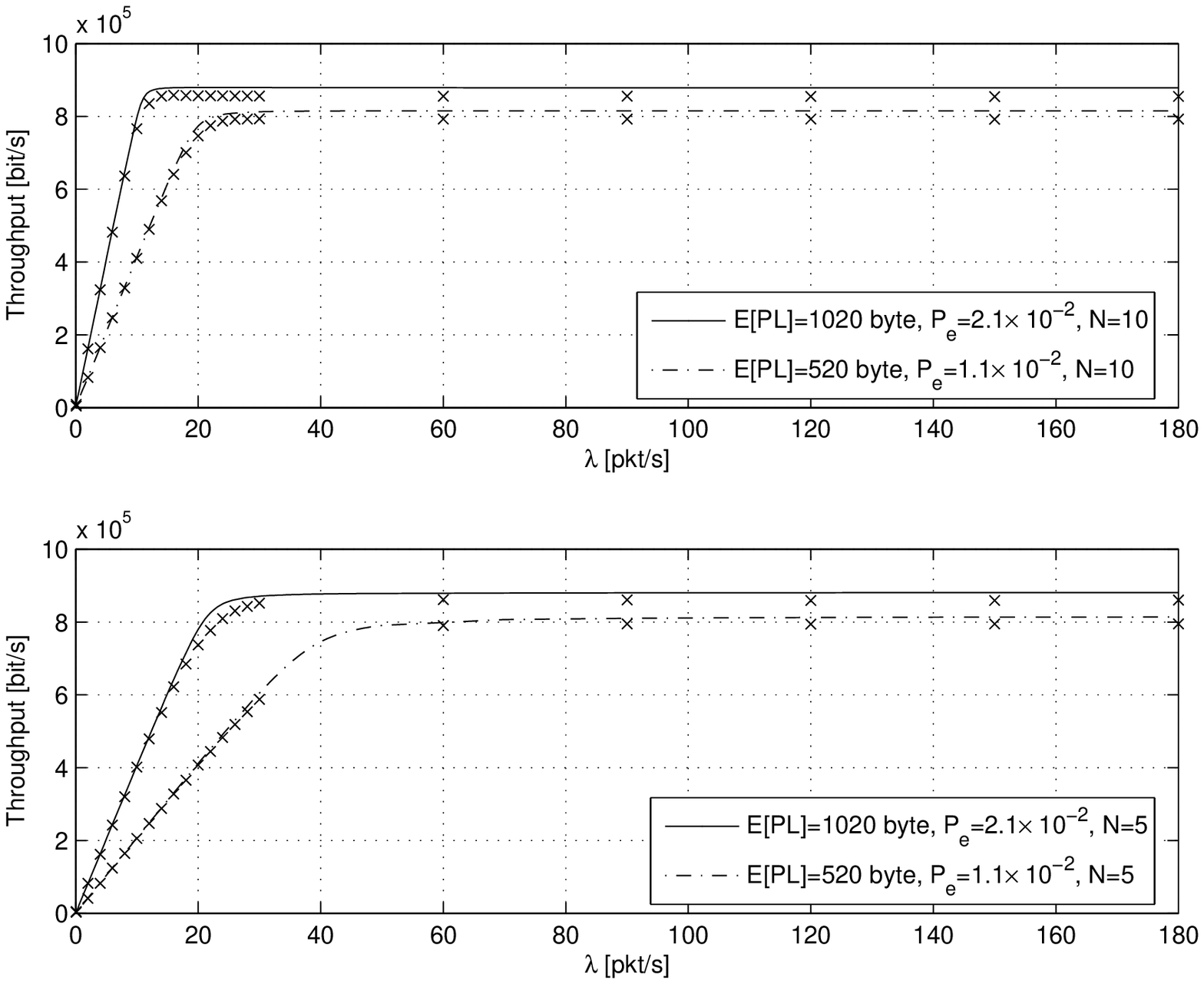}{Both subplots show
theoretical and simulated (ns-2) throughput for the 2-way
mechanism as a function of the packet rate $\lambda$, for two
different payload sizes and fixed packet error probability, as
noted in the legend. Capture threshold $z_0=1$dB in both subplots.
Simulated points are identified by cross-markers over the
respective theoretical curves. Upper subplot refers to a number of
contending stations equal to $N=10$, whereas the lower subplot
corresponds to $N=5$.}{N5_10_capture}
\section{Modelling the linear behavior of the throughput in unsaturated conditions}
The linear behavior of the throughput in unsaturated traffic
conditions, along with its dependence on some key network
parameters, can be better understood by analyzing the throughput
in~(\ref{eq.system2}) as $\lambda\rightarrow 0$, i.e., in unloaded
traffic conditions. From~(\ref{eq.tau}), it is easily seen that
the probability $\tau\rightarrow 0$ when $\lambda\rightarrow 0$,
since $q\rightarrow 0$. In particular, for very small values of
$\lambda$, the following approximation from~(\ref{eq.tau}) holds:
\[
\tau\simeq \frac{q}{1-P_e}
\]
In addition, it is straightforward to verify that
$P_{cap}\rightarrow 0$ (from (\ref{capture_probability})). As
$\lambda$ approaches zero, the throughput can be approximated by:
\begin{equation}\label{lin_model_throughput}
S\simeq \frac{N (1-P_e)E[PL]}{\sigma}\tau=\frac{N\cdot
E[PL]}{\sigma}q=N\cdot E[PL]\cdot\lambda
\end{equation}
since $q=1-e^{-\lambda E[S_{ts}]}$ can be approximated by $q\simeq
E[S_{ts}]\lambda$ upon employing MacLaurin approximation of the
exponential, and $E[S_{ts}]\rightarrow\sigma$ for
$\lambda\rightarrow 0$ as it can be deduced
from~(\ref{equ_E_Sts}).
The main conclusion that can be drawn from this equation is that
as $\lambda\rightarrow 0$, the throughput tends to manifest a
linear behavior relative to the packet rate $\lambda$ with a slope
depending on both the number of contending stations and the
average payload size $E[PL]$.

It is interesting to estimate the interval of validity
$[0,\lambda_c]$ of the linear throughput model proposed
in~(\ref{lin_model_throughput}). To this end, let us
rewrite~(\ref{eq.system2}) as follows:
\begin{equation}
\label{eq:throughput}
\begin{array}{rcl}
  S & = & \frac{E[PL]}{\left[T_s-\frac{T_c}{1-P_e}+\frac{T_e P_e}{1-P_e}\right] +
  \frac{\sigma\frac{1-P_t}{P_t} + T_c}{P_s(1-P_e)}} \\
\end{array}
\end{equation}
Since $T_s,T_c,T_e$ and $P_e$ are independent of $\tau$, $S$ can
be maximized by minimizing the function:
\begin{equation}
\label{eq:F_tau}
  F(\tau) = \frac{\sigma\frac{1-P_t}{P_t} + T_c}{P_s(1-P_e)}
\end{equation}
or equivalently, by maximizing the function:
\begin{table}\caption{Values of $\lambda_c$ from (\ref{lambda_critici}) for the
setup considered in Figs \ref{Throgh_lambda_par_z0_24dB}-\ref{Throgh_lambda_par_variN_z0}.}
\begin{center}
\begin{tabular}{|c|c|c|c|}\hline
\hline E[PL]=1024 byte&N=4&N=10&N=20 \\\hline

 SNR=20dB    & 13.9187 & 5.5372 &2.7638\\

SNR=45dB &   26.7546   &  10.6444 & 5.3132   \\

\hline\hline E[PL]=128 byte&N=4&N=10&N=20 \\\hline
SNR=45dB &   135.0307   &  53.3990 & 26.6039   \\

\hline\hline
\end{tabular}
 \label{lambda_c_simulazioni}
\end{center}
\end{table}
\begin{equation}
\label{eq:F_tau_inv}
  F_1(\tau) = \frac{1}{F(\tau)} = \frac{P_s(1-P_e)}{\sigma\frac{1-P_t}{P_t} + T_c}
\end{equation}
with respect to $\tau$. Upon substituting~(\ref{equat_Pt})
and~(\ref{equat_PS}) in $F_1(\tau)$, it is possible to write:
\begin{equation}\label{eq:F*}
    F_1(\tau) = \frac{N\tau (1-P_e)}{\sigma(1-\tau) + T_c(1-\tau)^{1-N} - T_c(1-\tau)}
\end{equation}
Differentiating $F_1(\tau)$ with respect to
$\tau$ and equating it to zero, it is possible
to write:
\begin{equation}\label{eq:de_F*_zero}
\begin{array}{rcl}
  (1-\tau)^N (\sigma - T_c) - T_c(N \tau - 1)                           & = & 0 \\
\end{array}
\end{equation}
Under the hypothesis $\tau \ll 1$, the following approximation
holds:
\[
(1-\tau)^N \approx 1 - N\tau + \frac{N(N-1)}{2}\tau^2 + o(\tau)
\]
By substituting the previous approximation
in~(\ref{eq:de_F*_zero}), it is possible to obtain the value of
$\tau$ maximizing $F_1(\tau)$:
\begin{equation}\label{eq:tau_max}
    \tau_m = \frac{\sigma-\sqrt{\sigma\left[N\sigma - 2(N-1)(\sigma-T_c)\right] / N} }{(N-1)(\sigma-T_c)}
\end{equation}
Substituting $\tau_m$ in~(\ref{eq:throughput}), it is possible to
obtain the maximum throughput $S_m$:
\begin{equation}\label{S_max_equat}
    S_m =  \frac{E[PL]}{\left[T_s-\frac{T_c}{1-P_e}+\frac{T_e P_e}{1-P_e}\right] + \frac{(\sigma-T_c)(1-\tau_m)^N + T_c}{N\tau_m(1-\tau_m)^{N-1}(1-P_e)}}
\end{equation}
Equating $S_m$ to the linear model
in~(\ref{lin_model_throughput}), it is possible to obtain the
value $\lambda_c$ for which the linear model reaches the beginning
of the saturation zone, i.e., the maximum $\lambda$ above which
the linear model loses validity:
\begin{equation}
\label{lambda_critici}
\begin{array}{rl}
  \lambda_c  = & \frac{1}{N\left[T_s-\frac{T_c}{1-P_e}+\frac{T_e P_e}{1-P_e}\right] + \frac{(\sigma-T_c)(1-\tau_m)^N + T_c}{\tau_m(1-\tau_m)^{N-1}(1-P_e)}} \\
\end{array}
\end{equation}
The linear model~(\ref{lin_model_throughput}) shows a close
agreement with both theoretical and simulated throughput curves.
Fig.~\ref{linearity_throughput_lambda} shows the straight
lines~(\ref{lin_model_throughput}) for three different values of
the number of contending stations $N$, compared with the
theoretical curves already depicted in
Fig.~\ref{Throgh_lambda_par_z0_24dB}. The figure also shows the
values of $S_m$ as derived from~(\ref{S_max_equat}) along with the
three values of $\lambda_c$ deduced from~(\ref{lambda_critici}).
Notice that $\lambda_c$, besides being the limit of validity of
the linear model, can also be interpreted as the value of packet
rate above which the throughput enters saturation conditions.

In order to further verify the values of $\lambda_c$ deduced from
(\ref{lambda_critici}), Table \ref{lambda_c_simulazioni} shows the
values of $\lambda_c$ related to both simulation and theoretical
results depicted in Figs
\ref{Throgh_lambda_par_z0_24dB}-\ref{Throgh_lambda_par_variN_z0}.
In particular, values shown in the upper part of the table, i.e.,
the ones labelled E[PL]=1024, are related to the curves in Fig.
\ref{Throgh_lambda_par_z0_24dB}, whereas the other values are
related to the results in Fig. \ref{Throgh_lambda_par_variN_z0}.
Notice the good agreement between theoretical and simulated
results even in the presence of capture. The key observation here
can be deduced by a comparative analysis of the results shown in
Figs.
\ref{Throgh_lambda_par_z0_6dB}-\ref{Throgh_lambda_par_z0_24dB}:
capture tends to simply shape the peak on which the throughput
reaches its maximum, leaving the abscissa of the maximum, that is
$\lambda_c$, in the same position.
\figura{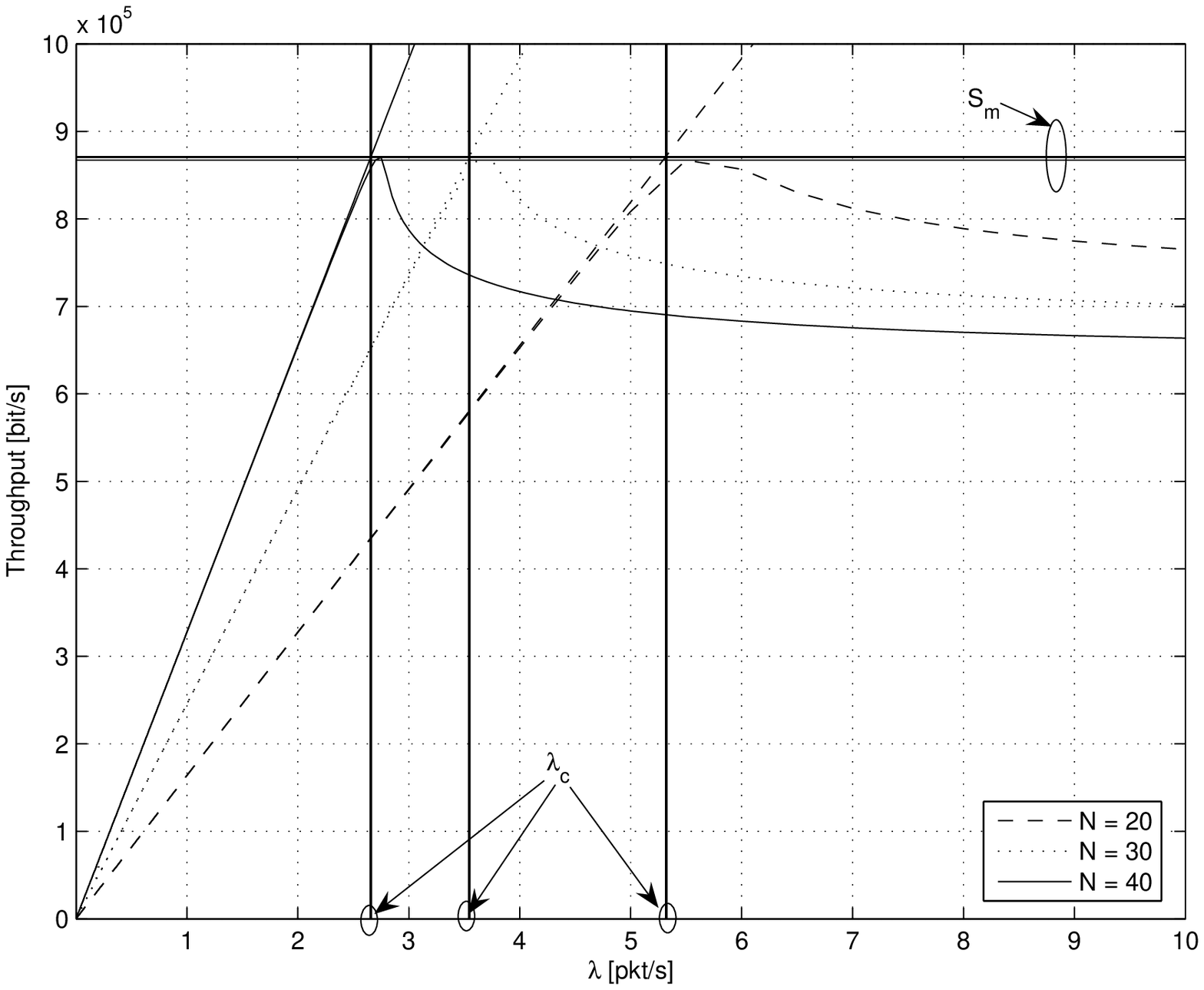}{Throughput for the
2-way mechanism as a function of the packet rate $\lambda$, for
three different number of contending stations. Capture threshold
and SNR are set to very high values since in pre-saturation
regime. Payload size is 1024 bytes. Straight lines refer to the
linear model of the throughput derived
in~(\ref{lin_model_throughput}).}{linearity_throughput_lambda}
\section{Conclusions}
In this paper, we have provided an extension of the Markov model
characterizing the DCF behavior at the MAC layer of the IEEE802.11
series of standards by accounting for channel induced errors and
capture effects typical of fading environments under unsaturated
traffic conditions. The modelling allows taking into consideration
the impact of channel contention in throughput analysis which is
often not considered or it is considered in a static mode by using
a mean contention period. Subsequently, based on justifiable
assumptions, the stationary probability of the Markov chain is
calculated to obtain the behavior of the throughput in both
unsaturated and saturated conditions. The closed form expressions
allow derivation of the throughput as a function of a multitude of
system level parameters including packet and header sizes for a
variety of applications. Simulation results confirm the validity
of the proposed theoretical models.
\end{document}